# Graph-Theoretical Description and Continuity Problems for Stress Propagation Through Complex Strut Lattices

Marcos A. Reyes-Martinez[1†], Alain Kadar[2, 3†], Steven Dunne[2], Sharon C. Glotzer[2], Christopher L. Soles[1,*] Nicholas A. Kotov[2,3*]

[1]Materials Science and Engineering Division, National Institute of Standards and Technology (NIST), Gaithersburg, 20899, MD, USA.
[2]Department of Chemical Engineering, University of Michigan, Ann Arbor, 48109, MI, USA.
[3]Center for Complex Particle Systems, University of Michigan, Ann Arbor, 48109, MI, USA.

*Corresponding author(s). E-mail(s): christopher.soles@nist.gov  kotov@umich.edu;
[†]These authors contributed equally to this work.

## ABSTRACT

Interconnected networks of rigid struts are critical for application in lightweight, load-bearing structures. However, accurately modeling stress distribution in these strut lattices poses significant computational challenges due to its strong dependence on organizational patterns, boundary conditions, and collective effects. Leveraging two-dimensional strut lattices that enable visualization of local elastic deformation, we investigate how graph theory (GT) provides a framework for stress prediction. We investigate how the geometric features often neglected by GT play a crucial role in the behavior of anisotropic networks. We also address the challenge of topological continuity that arises when applying discrete mathematics to physical structures. We show that modified centrality parameters combining lattice topology with geometry more accurately predict local stress, as validated through birefringence imaging and finite element modeling. Finally, we show how further improvements are made by incorporating strut lattice boundary conditions into the centrality definition, in a manner that simultaneously simplifies the computational cost.

## INTRODUCTION

Regular struts latices are ubiquitous in lightweight loadbearing structures. The emergence of metamaterials has highlighted the performance advantages of stochastic strut-based architectures for energy dissipation, long-term resilience, and reliable performance across diverse loading conditions.[1–4] Inspired by biological load-bearing structures, aperiodic strut lattices also provide the possibility of simultaneous optimization of multiple, often contradictory, mechanical properties. This unique capability ensures reliable performance under variable deformation directions as well as both periodic and static loads. However, even small changes in lattice geometry, strut connectivity, or material composition can lead to drastically altered mechanical responses in both regular and stochastic lattices.[5] These changes often result from phenomena such as shockwave resonances, nonlinear deformations, and long- and short-range collective effects,[6] which can culminate in catastrophic events like avalanche failures. Conventional computational methods for predicting stresses in complex structures, such as finite element models (FEMs), are often both time-intensive and resource-demanding, requiring numerical solutions for models that integrate rigid and soft elements. The inherent nonlinearity of deformations further complicates these analyses, creating intricate correlations between global stress propagation and localized failures. These challenges underscore the need for complementary approaches to support and enhance FEMs, especially for complex mechanical metastructures combing order and disorder.

Systems of interconnected rigid elements with virtually any degree of complexity can be described as *graphs, $\mathcal{G}(\mathcal{N}, \mathcal{E})$,* or sets of nodes and edges.[5,7,8] Graph Theory (GT) allows for predictions of electrical,[9–15] thermal,[11] and mechanical[16–20] properties using the notions of charge, heat, and stress transport through networks, which result in correlations between physical properties and several GT parameters.[21] For example, *betweenness centrality*[22,23] and other GT parameters[7] characterizing topological distances and connection patterns[24] can identify struts where the mechanical stress or electrical current are concentrated at a fraction of the computational cost required for FEMs.[2,20,25] Furthermore, Schrödinger or wave equations solved on so called quantum graphs[26] can describe stress flow through the network,[27] which can incorporate the methods of quantum chemistry/physics to further facilitate the computations. Importantly, GT methods are applicable to a wide range of loadbearing structures inclusive of networks from spheroidal particles,[20,23,28,29] auxetic lattices,[30] and biomimetic composites.[31]

Although GT may offer highly efficient coarse-grainings for material design, geometric features of material networks are usually lost with these approaches. And whilst the need to incorporate geometry into GT descriptions of physical networks has been acknowledged,[18,32,33] there are scientific problems especially pertinent to material systems that must still be addressed. One of them is the problem of continuity of graph representations. It emerges when methods of discrete mathematics are applied to strut lattices with gradually variable geometric parameters. Then, seemingly identical methodologies for mapping strut lattices onto graphs can lead to reasonable predictions for some structures and topological parameters but not for others.[7,27,28] Another underexplored problem is that behavior depends on boundary conditions, which has direct implications for stress predictions and hence for thresholds of local failures.

This study addresses these fundamental aspects of graph-structure relations for both regular and stochastic lattices by introducing (**1**) variable node-edge representation of struts accounting for local geometries; (**2**) structural descriptors combining topological and physical characteristics and (**3**) targeted selection of shortest paths contributing to betweenness centrality near boundaries. The overall methodology was validated for different configurations of strut lattices using birefringence videography (**Figure 1a**) and FEM simulations.

**RESULTS**

**Experimental models and their graphs**

In this work, regular and stochastic strut lattices in physical space will be referred to as *networks*. The key difference between a *network* and a *graph* is that the latter is a mathematical abstraction of the former in topological space. To explore graph-structure relations we employed two-dimensional (2D) macroscale networks with both regular and stochastic patterns laser-cut from transparent sheets of poly(methylmethacrylate) **(Figure 1b)**. The well-known strong photoelastic behavior of poly(methylmethacrylate)[23] enables visualization of local stress using polarized light. Regular networks in this study include rhombic lattices with angle of strut intersection of 90° as well as square (4,6,12) Archimedean lattices. The stochastic networks include auxetic lattices, generated using an algorithm by Reid et al[30] as well as networks replicating patterns of self-assembled aramid nanofibers (ANFs).[8]

We found that adequate representations of the networks can be surprisingly nonobvious even for relatively simple 2D networks. It might be intuitive to expect that there is a one-to-one mapping of a 2D network onto a graph when struts are represented as edges and their intersections are represented as nodes. However, such models might not be the best for stochastic networks where struts can intersect with a wide range of angles or even through families of ordered lattices with different geometric variations of fixed topological patterns. For example, imagine a pin-jointed rhombic lattice section with struts having a non-zero, fixed thickness (**Figure 1c**). Its internal angle, $\theta$, determines the characteristics of the strut intersection. Let us assume that for $\theta = 90°$ each strut intersection is represented by a single node. When $\theta$ is small (~15-20°), the strut intersection may no longer be represented by a single node, requiring a short edge (and hence extra node) to represent it (**Figure 1d**). For $0° < \theta < 10°$, the edge representing the intersection becomes approximately the same size as the edges representing struts (**Figure 1e**). If the struts are thick enough, even a rhombic network with $\theta = 90°$ will exhibit this unintuitive behavior. To extract a graph from the network in the most unbiased and reproducible manner we use our open-source package, *StructuralGT*, without merging any nodes (details in Methods).

**Relating stress and structure for strut lattices**

Multiple GT parameters can be used to describe strut lattice structures. These parameters include *average degree*, which is the average number of edges connected to each node; *average clustering coefficient*, which is higher when it is more likely for connected nodes to share common neighbors; *assortativity coefficient*, which is higher than zero when similar nodes are connected to each other, and lower when dissimilar nodes are connected; the *average nodal connectivity*, which is the average number of node removals required to disconnect a randomly selected pair of nodes. *Edge length deviation* gives a comparison of how the length of edges vary within each network (and is defined as the ratio of the edge length standard deviation to mean). While each individual parameter cannot unequivocally describe a network, their set distinguishes the **structure** of strut lattices in short-, medium- and long-ranges.

Extending this GT framework to distinguishing different **properties**, we may develop specialized parameters that combine network topology, geometry, and physical parameters that can be even more specific than topological sets. For pedagogical purposes, let us first consider the GT parameter *geodesic edge betweenness centrality* ($EBC_G$) defined for edge $e$ as

$$EBC_G(e) = \frac{1}{2N(N-1)} \sum_{s,t \in \mathcal{N}} \frac{\sigma_{st}(e)}{\sigma_{st}} \qquad 1$$

where $e$ refers to a specific edge, $N$ is the number of nodes. $\sigma_{st}$ is the number of shortest paths between a given source and target, $s$, $t$, respectively. $\sigma_{st}(e)$ is the number of shortest paths between $s$, $t$ containing $e$. Typically, the "shortest path" in the context of GT refers to the shortest sequence of adjacent edges that connects two nodes. Similarly to other centrality measures,[7,22–24] $EBC_G$ can be used to identify bottlenecks for transport through networks of current, heat, and stress.[34]

The aforementioned problem of continuity clearly presents itself when considering a common situation of nodes for intersections with small $\theta$ angles, where the struts can also be curved. Such intersections can be represented by one, two, or even more nodes **(Figure 1c-e)**, which has been pointed out in the past for networks of nanofibers.[35] In the case of data sets containing many thousands of nodes, one may deliberately merge the closely positioned nodes to simplify the analysis, but the conceptual problem of the mismatch between a network in physical space and its abstraction as a graph in topological space would nevertheless remain. As such, graph representations go through discontinuous changes when nodes are added or removed. The network-graph discontinuity is exacerbated when the struts have thicknesses of similar magnitudes to the voids. In all these cases, the point at which to assign two nodes to a strut is arbitrary.

In other words, minor structural variations in the network manifest as inadequately abrupt changes in the parameter being used for critical assessment of the local flow of current, heat, and stress and thus, reliability of the structure. We therefore argue that GT parameters must depend on continuous material changes to adequately map the continuity of the geometry.

With geometric and topological characteristics of the lattice being important for their mechanical properties, we weight the edges by their length, which also contributes to the continuity of the GT representation of the lattice. The resulting definition of shortest path corresponds to the sequence of adjacent edges connecting two nodes with minimal total Euclidean distance travelled along the edges comprising the list.

The appropriate terminology for a betweenness centrality weighted by length, is to substitute "geodesic" for "length weighted". In **Figure 2b** and **c**, we use a simple network to compare $EBC_G$ with *length-weighted edge betweenness centrality* ($EBC_L$), respectively. While all edges are equal for $EBC_G$, $EBC_L$ distinguishes between the edges based on their length, which is certainly essential for stress transfer. We note that for networks with edges of approximately equal length (e.g., those from packings of equally sized particles), length weighting is less important, and $EBC_G$ and $EBC_L$ perform similarly.

While $EBC_G$ and $EBC_L$ capture organization of the network on a global scale because of the summation of all the short paths through each edge, they are invariant to boundary conditions, which is problematic for an array of properties and external stimuli (e.g. direction of uniaxial compression). To show how this conflicts with our intuition, we calculate $EBC_G$ for a square lattice unit cell (or $K_{1,4}$ complete graph) in **Figure 2d**. In the $K_{1,4}$ graph, every edge has the same value of $EBC_G$. But under uniaxial compression, one would expect stress to concentrate along the edges aligned with the direction of compression. Therefore, our parameter should also be some function of the boundary conditions applied to the material. To incorporate this, instead of summing over all source-target pairs (as in 1), we select the subset of source-target pairs in contact with the perturbed boundaries and call our parameter *edge boundary betweenness centrality* ($EBC_B$). In this definition, we designate all nodes in contact with the source of stress as source nodes. All nodes along the opposite plane are designated as target nodes. The use of length weighting gives the parameter a "*length weighted*" prefix, denoted $EBC_{LB}$, for simplicity:

$$EBC_{LB} = \frac{1}{2T(S-1)} \sum_{s \in \mathcal{S}, t \in \mathcal{T}} \frac{\sigma_{st}^w(e)}{\sigma_{st}^w} \qquad 2$$

where $S$ and $T$ are the number of source and target nodes $\mathcal{S}$ and $\mathcal{T}$ are sets of source and target nodes, respectively. $\sigma_{st}^w(e)$ is the number of length-weighted shortest paths between $s$ and $t$ containing $e$. The marked difference between $EBC_G$ with $EBC_{LB}$ can easily appreciated in **Figures 2d** and **2e**.

To further highlight the difference between $EBC_G$ with $EBC_{LB}$, we compare the computation time for both, across a series of samples obtained from birefringence and finite element analysis (FEA, **Figure 2f**). Because the modification we use here reduces the number of pairs of vertices between which shortest paths must be found from $\sim N^2$ to $\sim ST$, the scaling for number of shortest path calculations required is reduced approximately by a factor of $N$ for square samples. Even after accounting for the more expensive weighted shortest path algorithm (Dijkstra instead of breadth first search), $EBC_{LB}$ remains favorable to $EBC_G$, scaling as $O(n^{1.5}log(n))$, as opposed to $O(n^2)$ for $EBC_G$. Details for scaling and benchmarking calculations are given in the SI.

While we arrived at this form of $EBC_{LB}$ without imposing any constitutive relations or any other mechanical characteristics on the networks, the results agree with the classical theories of beams. Definitions and analysis of length-weighting has a direct connection to the definition of axial forces in strut-based lattices. For example, the axial force, $P$ of a pin-jointed member of the network is $P \sim EA/l$, where $E$ is the Young's modulus, $A$ is the cross-sectional area and $l$ is the length of the strut.[4,36] This connection can be investigated numerically for virtually any structure using FEA. Consider the kite structure in **Figure 3a**, subjected to a uniaxial compressive stress, $S_T$, applied on the top as indicated. The stresses to the left and right side of the top of the structure are $S_l$ and $S_r$, respectively. To illustrate how strut length affects stress distribution in a structure in static equilibrium we performed FEA of the structure, and plotted the ratio of average axial stresses in the right and left struts, $S_r/S_l$, as a function of increasing elongation of one half of the kite, $l/l_0$ (**Figure 3b**). As the initial length of the two struts on the right side of the structure increases, the magnitude of the stresses on the right side rapidly decreases. **Figure 3b** also shows the ratio of GT parameters, $EBC_{LB}$ and $EBC_G$, for the left and right struts. A perfect GT parameter would be identical to the FEA $S_r/S_l$ curve. As expected, $EBC_G$ does not change as a function of $l/l_0$. Although $EBC_{LB}$ does not follow $S_r/S_l$ exactly, it provides a step-function approximation to the nearly discontinuous redistribution of stress.

While we have discussed the benefits of mitigating of discreteness of GT parameters (**Figure 3**), modification of $EBC_{LB}$ to smooth the discontinuity also has notable disadvantages, elaborated in the SI. An analytical solution for the axial stresses of a similar family of structures under load is developed in the SI and the results give similar trends to **Figure 3**.

**Predictions of stressed edges**

Following Berthier et al.[20] the relative magnitude of an edge's GT parameter can be used to predict its response to stress. Stressed edges can be classified for an array of different network structures (a1 through d1, **Figure 4**). The edges are expected to carry greater stress when $EBC_{LB} > \langle EBC_{LB} \rangle$, where $\langle . \rangle$ is the average parameter value over all edges within a strut lattice. In the middle column of **Figure 4** (a2 through d2), edges with $EBC_{LB} > \langle EBC_{LB} \rangle$ are highlighted in yellow. $\langle EBC \rangle$ can be adapted to each lattice: when the stress classification cutoff for $\langle EBC \rangle$ is decreased, more edges are classified as stressed, which is important for understanding the likelihood of consequential misclassification and potential edge failure. Overall, the impact of the cutoff values on classification can be compiled as a receiver operating characteristic (ROC) curve to assess overall performance.[30] The greater the area under the curve, the better the performance (**Figures 4,** a3 through d3). In all cases, $EBC_{LB}$ outperforms $EBC_G$, because of inclusion of the boundary conditions and length weighting. The performance discrepancy increases from **Figure 4a3** to **Figure 4d3**, as the strut lattices become increasingly ordered and, consequently, anisotropic. Therefore, the inclusion of the boundary conditions becomes increasingly important.

For example, the network in **Figure 4c1** is an (4,6,12) Archimedean lattice with a discrete *edge orientation distribution* (EOD, Figure S4c). To demonstrate the necessity of considering boundary conditions in this instance, we note that, in the absence of geometric considerations, a strut in the infinite (4,6,12) Archimedean lattice can be in one of only three environments. However, some edges that are topologically equivalent (or degenerate) will be in different geometric environments under uniaxial compression because they will be more aligned with the direction of compression, and so will experience different amounts of stress, thus lifting the degeneracy. Furthermore, every edge in the rhombic lattice is topologically ***and*** geometrically equivalent when the direction of compression is aligned the vertical axis (**Figure 4d1**). For a finite sample, their degeneracy is lifted by their proximity to the lattice boundary. In both the stress and $EBC_{LB}$, this simplicity and order causes high values to predictably distribute themselves in the shape of an hourglass (**Figure 4d2**).

In prior works, we showed that ANF nanofibers (**Figure 4a1**) can self-assemble into stochastic networks with uniform EOD (Figure S4a). But because the other networks in this study were generated in a less random manner, their EODs are less uniform which translates into a more anisotropic mechanical response. For example, the network in **Figure 4b1** was obtained using an algorithm for generating complex auxetic lattices.[30] By calculating the EOD, we see this algorithm aligns the majority of edges with the reference axes of the image (i.e. 0∘ and 90∘ (Figure S4b). Because of this, the edges under stress strongly depend on the direction of compression. **Figures 5a, b** shows how the stressed edges change when the sample is compressed in different directions. To obtain a prediction of stressed edges from graph models, the GT parameter should be adjusted according to this boundary condition. **Figures 5e, f** show how $EBC_{LB}$ redistributes itself across stressed edges when the source and target nodes are changed, because of the different direction of compression. The value of $EBC_G$ remains similar for two copies of the same network  one cut horizontally, one vertically (**Figures 5c** and **d**). The sensitivity and accuracy of this analysis of stress distribution could be assessed from the fact that the high $EBC_G$ edges are not identical between **Figures 5c** and **Figures 5d**. Compressing in a different direction requires cutting a different sample. Albeit following the same geometrical model, the samples have minor differences, which is picked up by the image analysis.

Figure S5 shows how this sample cutting effect is removed by carrying out FEA for networks from ANF structures, confirming both accuracy and sensitivity of the GT analysis in this framework.

Finally, we note the distinction between our approach and spring networks.[37] While both can be used for predicting mechanics in complex systems, spring network models are mathematically equivalent to a modified form of $EBC_G$, which is discussed in the SI, and shown to fail in predicting accurate distributions of stress in strut lattices.

**DISCUSSION**

Here we elaborate the problem of adequacy of graph models and GT parameters for predicting stress distribution in architected strut lattices based on the notions of percolation[29] and flow[38] of stress through complex networks. An essential requirement for such models is continuity of GT parameters for determining stress concentration points, which conflict with the framework of discrete mathematics, while being essential for material design. It can be addressed by the adaptive placement of nodes in graph models using dedicated software packages, such as *StructuralGT*.[39] This task also requires utilization of parameters combining topological, geometric, and physical characteristics of materials. From a practical perspective, it is important to point out that the incorporation of boundary conditions can be most efficiently achieved by *removing* computations from the parameter definition, as opposed to adding them. This study shows that the removal of less important shortest paths for betweenness centrality, reduces the number of required computations by an order of magnitude, while simultaneously improving the classification of stressed edges.

Two limitations of the graph models should also be noted (see details in SI). First, the uniformity of regular lattices make them more susceptible to experimental error and implicit, but often incorrect, assumptions exemplified by the uniformity of the applied loads. Such assumptions could be easily enshrined in the GT models, which directly impacts the accuracy of stress predictions. Second, this approach is particularly suitable for the design of materials with high void fractions being optimized for lightweight and, for instance, Young's modulus. The GT methodology becomes progressively less applicable for strut lattices with gradually decreasing void fraction. For example, 'lattices' with thick struts start behaving like a continuous sheet and thus, should not be represented as a graph.

There are several avenues which this work can be taken further. The use of mechanochemically responsive polymers for deformation visualization[40] leaves extension of the analysis from 2D to 3D networks the most obvious future pathway. Additionally, the development of material and time dependent parameters, exemplified by $EBC_{LB}(t)$, will be helpful for propagation of waves in strut lattices, which is of great practical importance.

## METHODS

Certain instruments and materials are identified in this paper to specify the experimental details adequately. Such identification does not imply a recommendation by the National Institute of Standards and Technology or that the materials are necessarily the best available for the purpose.

**Sample fabrication.** Network samples were laser cut from 1.5 mm thick acrylic sheets (McMaster-Carr), using a Full Spectrum Laser, Muse 3D laser cutter.

**Mechanical Measurements.** Quasi-static, uniaxial compression experiments were performed on acrylic network structures using a Stable Microsystems Texture Analyzer TA, XTplus. Experiments were displacement-controlled at a rate of 0.1 mm s$^{-1}$ for both loading and unloading. The sample was illuminated using linearly polarized light and a cross-polarized digital camera (JAI BM-500GE) captured images of the deformed sample.

**Computational details.** Extraction of the graphs from images of experimental samples and FE was carried out with our package *StructuralGT*.[39] Calculation of betweenness centrality parameters were also carried out using functions that are part of this package.

**Supplementary information.** Accompanying supplementary files available online.

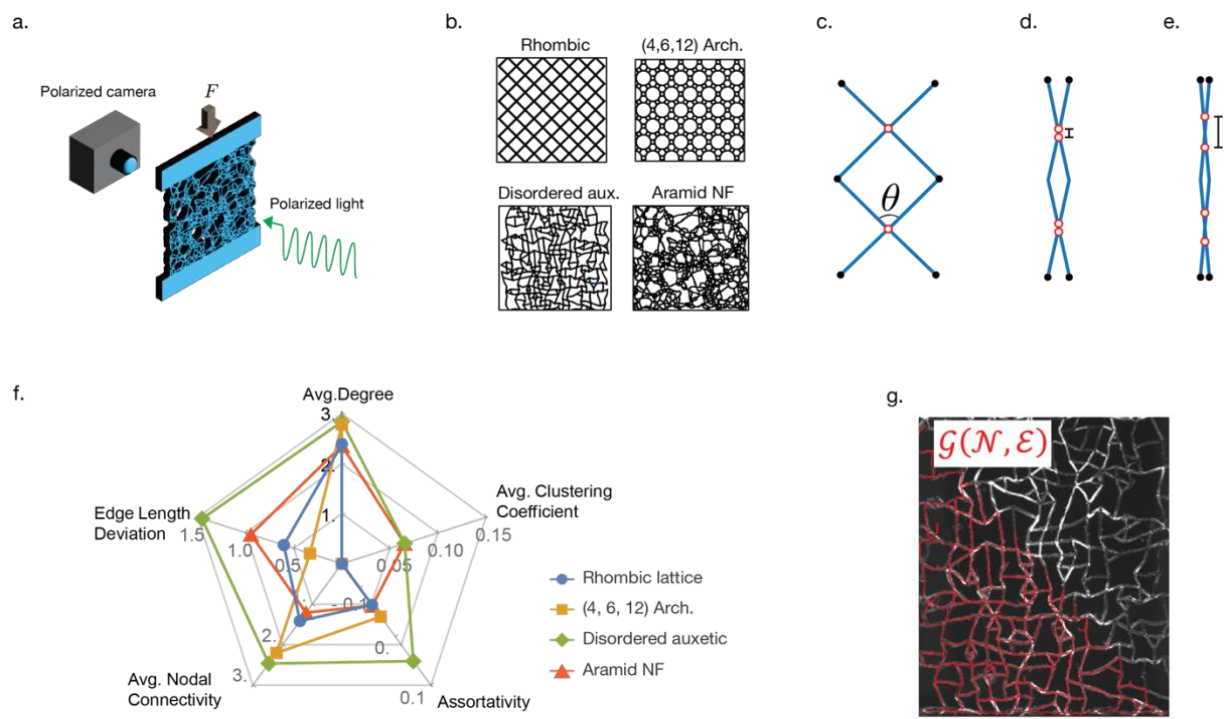

**Figure 1. Visualizing stresses in 2D networks.** (a) Laser-cut acrylic network samples are subjected to uniaxial compression. Stressed struts are captured by polarized photography. (b) The different type of network structures tested include ordered lattices such as rhombic and (4, 6, 12) Archimedean lattices, a complex auxetic metamaterials generated computationally and complex fiber network extracted from a micrograph of an aramid nanofiber sample. (c-e)

Depiction of network discontinuities in rhombic lattices with different angles of strut intersection, $\theta$, showing that when becomes increasingly small, single strut intersections (red ovals) are represented by multiple nodes in their graph abstraction. Without length weighting, the GT representation of (d) and (e) are identical. (f) Radial plot comparing the GT descriptors for networks in b. The different values across the parameter space highlights the differences in the graphs describing the physical networks. (g) Polarized photograph of 2D complex auxetic network while stressed. The brightness intensity of individual struts is directly proportional to the magnitude of the stress experienced by the strut. The graph, $\mathcal{G}(\mathcal{N}, \mathcal{E})$, corresponding to the 2D auxetic network sample is shown in red.

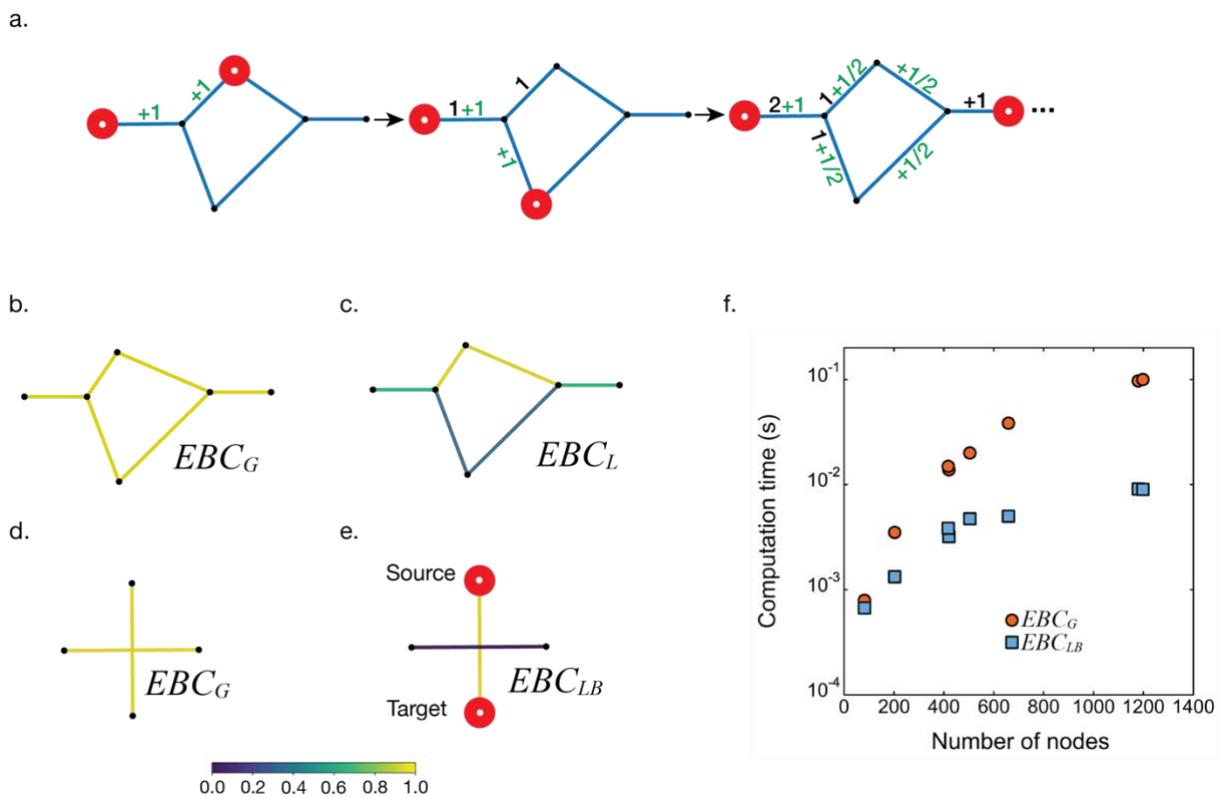

**Figure 2. GT parameters for strut lattices.** (a) Calculation of *geodesic edge betweenness centrality* ($EBC_G$). On each iteration, the algorithm identifies the edges comprising the shortest

path between the highlighted pair of nodes (red). The $EBC_G$ values of these edges are increased according to Eq. 1 (green). (b-c) Comparison between unweighted and weighted betweenness. Weighting allows distinction between short and long paths. (d-e) Comparison between $EBC_G$ and *length-weighted edge betweenness centrality* ($EBC_L$) for the $K_{1,4}$ graph. (f) Comparison of computation time as a function of number of nodes for both $EBC_G$ and $EBC_{LB}$ parameters.

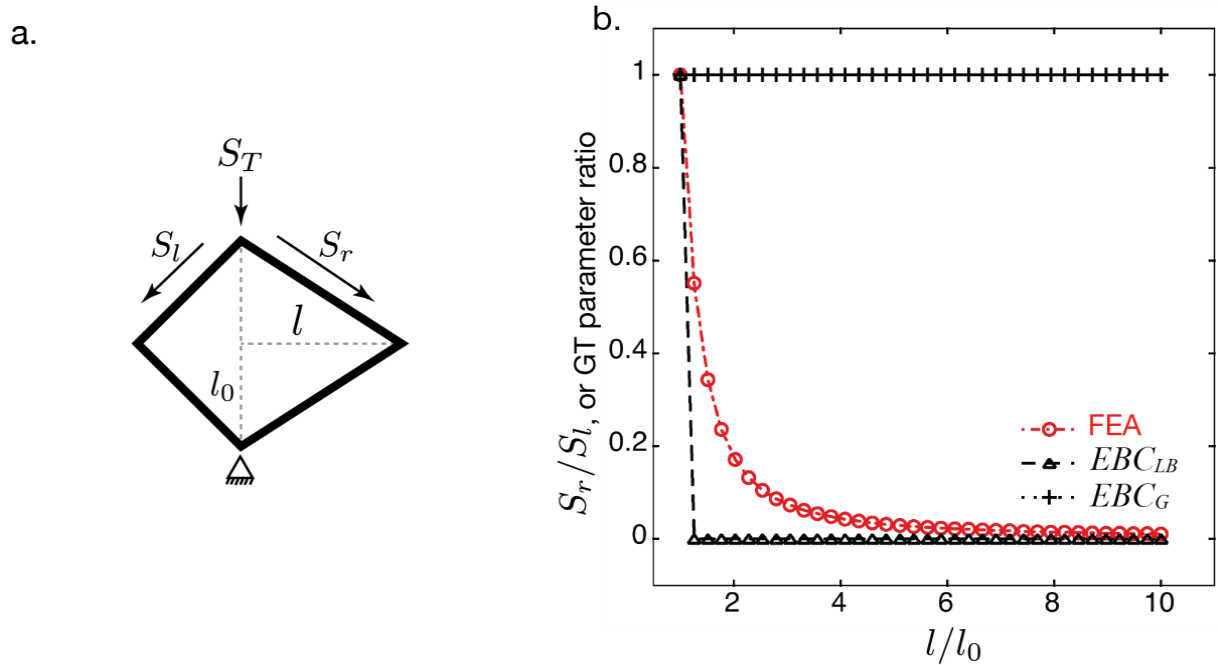

**Figure 3. Comparison of the finite element analysis (FEA) and GT parameters for the kite under uniaxial compression. (**a) Example of rigid-jointed struts forming a rhombus. Uniaxial compressive stress, $S_T$, imposed on the top. Stresses, $S_l$ and $S_l$ represent the stresses experienced by the left and right members of the rhomboid structure, respectively. The vertical dimension of the structure is $2l_0$, and the horizontal, $l_0 + l$. The graph (b) shows the ratio $S_r/S_l$ and the ratio between values of GT parameters in the left and right members: $EBC_G$ and $EBC_{LB}$.

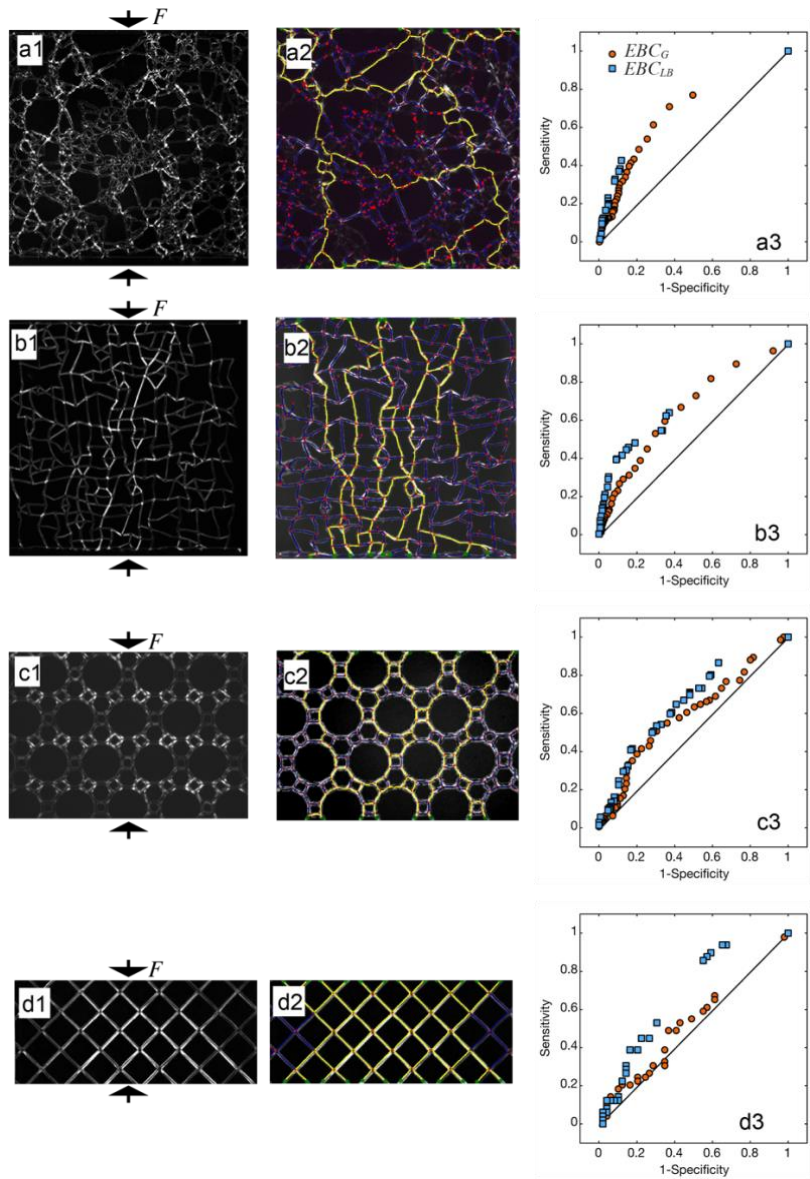

**Figure 4. Stress predictions based on $EBC$.** The first column (a1-d1) displays birefringence images of laser cut samples while being uniaxially compressed. (a1) Macroscale network laser cut from poly(methylmetacrylate) sheets with 2D architecture extracted from the 2D projection of aramid nanofiber network imaged by scanning electron microscopy. b1) Complex auxetic network generated from the pruning of struts from a sphere packing algorithm (c1) (4, 6, 12) Archimedean lattice and (d1) corresponds to 90° diamond lattice. The prediction of struts experiencing high stress are shown in the second column (a2-d2) highlighted in yellow over experimental photographs. The third column (a3-d3) shows ROC curves comparing predictions of $EBC_G$ and $EBC_{LB}$ for each of the tested network structures. Predictions are improved for $EBC_{LB}$ for all cases. The graph models were extracted using *StructuralGT*. No node merger or pruning options were applied.

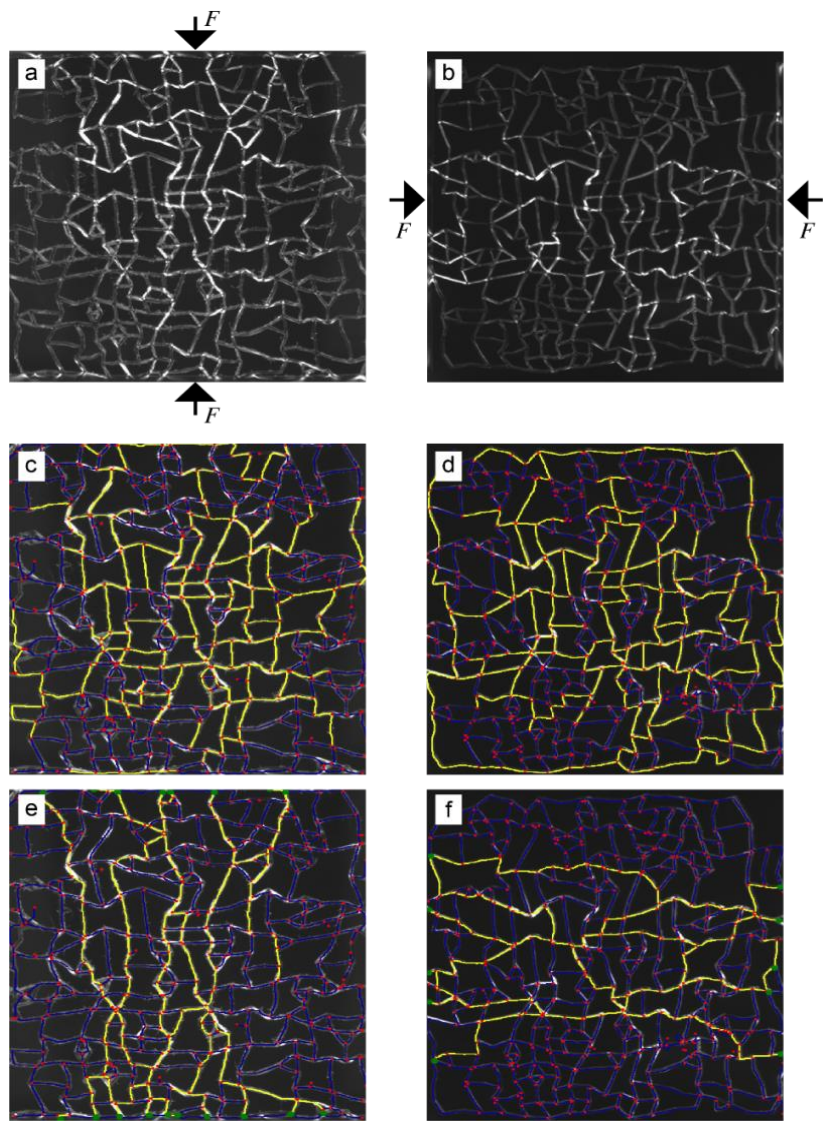

**Figure 5. Effect of changing boundary conditions on betweenness parameters for the complex auxetic lattice**. Left column is under vertical compression, right column is under horizontal compression. (a,b) show the raw image, (c,d) show the image with high $EBC_G$ edges highlighted, (e,f) show the image with high $EBC_{LB}$ highlighted.